# Mathematical Modeling of Arterial Blood Pressure Using Photo-Plethysmography Signal in Breath-hold Maneuver

Armin Soltan zadi, Raichel M. Alex, Rong Zhang, Donald E. Watenpaugh, Khosrow Behbehani

*Abstract—* recent research has shown that each apnea episode results in a significant rise in the beat-to-beat blood pressure and by a drop to the pre-episode levels when patient resumes normal breathing. While the physiological implications of these repetitive and significant oscillations are still unknown, it is of interest to quantify them. Since current array of instruments deployed for polysomnography studies does not include beat-to-beat measurement of blood pressure, but includes oximetry, it is both of clinical interest to estimate the magnitude of BP oscillations from the photoplethysmography (PPG) signal that is readily available from sleep lab oximeters. We have investigated a new method for continuous estimation of systolic (SBP), diastolic (DBP), and mean (MBP) blood pressure waveforms from PPG. Peaks and troughs of PPG waveform are used as input to a 5$^{th}$ order autoregressive moving average model to construct estimates of SBP, DBP, and MBP waveforms. Since breath hold maneuvers are shown to simulate apnea episodes faithfully, we evaluated the performance of the proposed method in 7 subjects (4 F; 32±4 yrs., BMI 24.57±3.87 kg/m$^2$) in supine position doing 5 breath maneuvers with 90s of normal breathing between them. The modeling error ranges were (all units are in mmHg) -0.88±4.87 to -2.19±5.73 (SBP); 0.29±2.39 to -0.97±3.83 (DBP); and -0.42±2.64 to -1.17±3.82 (MBP). The cross validation error ranges were 0.28±6.45 to -1.74±6.55 (SBP); 0.09±3.37 to -0.97±3.67 (DBP); and 0.33±4.34 to -0.87±4.42 (MBP). The level of estimation error in, as measured by the root mean squared of the model residuals, was less than 7 mmHg.

## I. INTRODUCTION

Recently, investigators have shown the presence of significant oscillations in nocturnal blood pressure (BP) in a sample of obstructive sleep apnea (OSA) patients [1]. The significant level of these oscillations and their high frequency (occurring at every apnea episode) have heightened the interest of researchers to explore the extend of these oscillations in the larger samples of OSA patients, as such investigation may provide clues to root cause of the cardiovascular and cerebrovascular comorbidities of OSA. The knowledge of the extend of these oscillations is particularly important because numerous studies have shown sufficiency of the systolic blood pressure (SBP) and diastolic blood pressure (DBP) and computed Mean arterial blood pressure (MBP) as a predictor of mortality [2-5].

With current state of the art technology, quantification of the nocturnal oscillations in BP requires the use of rather costly devices that can provide beat-to-beat measure of BP (e.g., Finapres) in sleep studies. Considering that larger population studies for this purpose, by necessity, involves multiple sleep labs. Given that measurement of nocturnal beat-to-beat BP is not part of the polysomnography studies in the sleep labs, it is of interest to develop methods of estimating the extent of apnea-driven BP oscillation from the equipment that are currently part of the standard polysomnography instrumentation.

Recent studies have highlighted the benefits of using photoplethysmography (PPG) signal that is available from most, if not all, sleep laboratory oximeter to estimate the BP oscillations indirectly [6]. Indeed, some investigators have explored using pulse transit time (PTT) extracted from PPG to estimate BP in awake subjects [7-9]. These studies have shown a correlation coefficient of around 0.8 in experimental studies. However, the estimation of nocturnal BP oscillation has not been studied before.

The objective of this study is to estimate the key features of blood pressure from PPG. Specifically, we estimate systolic BP (SBP), diastolic BP (DBP) and mean arterial BP (MBP) from PPG peaks and troughs. The motivation is to accurately estimate and track the changes of BP during simulated sleep apnea by using photoplethysmograph probe which is easier to apply and less costly.

## II. METHOD

We used peaks and troughs of the PPG waveform to model SBP and DBP correspondingly using Autoregressive Moving-average (ARMA) models. Using the estimated values of systolic and diastolic BP, we then estimated the MBP. ARMA model is a system identification method which can provide a mathematical model for the dynamics of the system as well as any pure time delay [10]. It has been shown to be efficient for the biological systems with multiple inputs and delays [11].

For the purposes of this study, we apply a single input (i.e., PPG feature) and single output (i.e., BP feature). A single-input-single-output, time-invariant, and causal ARMA model can be represented as a dynamic difference equation involving present and past values of the input and output as described in equation (2):

$$y(m) + a_1 y(m-1) + \cdots + a_{n_a} y(m-n_a) =$$
$$b_1 u(m-n_k) + \cdots + b_{n_b} u(m-n_b-n_k+1) + e(m) \quad (1)$$

where m is the sample number, y is the output, u is the input

Soltan zadi, K Behbehani, D. E. Watenpaugh are with the University of Texas at Arlington, Arlington, TX 76019 (armin.soltanzadi@uta.edu, and kb@uta.edu, ). Raichel M. Alex, 80523 USA (raichelmary.alex@mavs.uta.edu ). Rong Zhang is with the Neurology Department, University of Texas South Western, Dallas, TX 75390 USA (e-mail: RongZhang@texashealth.org).

and e stands for error, $a_i$ for $i = 1,2 ... , n_a$ and $b_j$ for $j = 1,2, ... , n_b$ are the model parameters that need to be computed, and $n_a$ and $n_b$ signify the orders of ARMA model while $n_k$ is the number of pure-time delay samples. A major decision in developing an effective ARMA model is the selection of the order of the model and any pure-time delay that may be involved. In this study, we use principle of parsimony and model adequacy to find the least mean square error in a certain range of orders. A compromise was reached for the orders and the delays to be 5. Then ARMA model parameters are estimated in MATLAB using least square method to minimize the model error for specific orders. All the model subsets (with all possible combinations of orders) are generated and compared by their MSE. The lowest MSE is selected to be best model generated in the specified $n_a$ and $n_b$ and $n_k$. There are other methods including Akaike's information criterion (AIC) [12-13] to find the best model but it does not necessarily give the lowest MSE. In our study, we limited the model orders to be less than 5 and therefore there is no need for finding simplest model. These orders are selected based on trial and error as we compared higher orders models and did not see any significant improvement in results.

### A. Focal points on BP signal

For this study, we opted to use SBP, DBP and MBP since these are proven to be in wide use in clinical practice as robust indicators of blood pressure health. The measured MBP was obtained from measured SBP and measured DBP using Eq. (2):

$$MBP = \frac{2DBP+SBP}{3} \qquad (2)$$

In estimating the model parameters for estimating SBP and DBP (Eq. (1)), we use measured SBP and DBP as outputs and peaks and troughs of PPG signal as inputs, respectively. From estimated SBP and DBP – denoted as $\widehat{SBP}$ and $\widehat{DBP}$ –, we compute the estimated MBP (i.e., $\widehat{MBP}$ using Eq. (2). Since the application of the ARMA model (Eq. (2)) requires equidistance sampling of both the input and output data, we use cubic spline method to interpolate DBP, SBP and MBP values at the same sampling rate of the BP and PPG signal, i.e., 100 Hz.

### B. Experimental setup

The protocol and written subject consent form for testing subjects were approved by our Human Subject Institutional Review Board. Seven subjects (4 F, 32±4 yrs., BMI 24.57±3.87 kg/m²) with no known ailment volunteered for this study and signed the consent form. The subjects were asked to avoid any caffeine intake for 6 hours before the experiments. The subjects were tested in supine position, performing a sequence of five breath holding maneuvers. The sequence of the maneuvers is shown in Figure 1. At the start, each subject was asked to breathe normally for 60 s to obtain baseline data. Afterward each subject performed a series of five breath hold maneuvers to induce dynamic changes in BP change. During each breath hold (BH), subjects were instructed to hold their breath for as long as they can. Hence, the duration of each breath hold varied, depending on the ability of the subject to prolong the breath hold. Inter breath hold intervals, i.e. normal breathings (NB), were fixed at 90 seconds to provide adequate recovery time between consecutive breath holds. Data was collected for the entire duration, from the initial baseline through the final 'ending NB' period (Fig. 1).

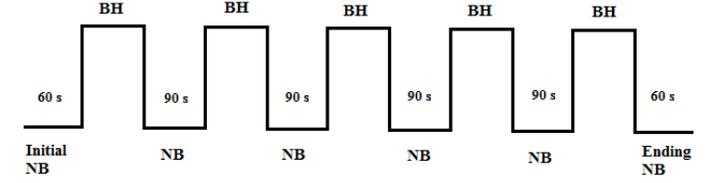

Figure 1- Timing diagram of protocols (NB: Normal breathing, BH: Breath hold)

We used a Finapres blood pressure monitor (Finapres Medical Systems, Enschede, Netherlands) for beat-to-beat measurement of BP [14]. Photoplethysmography signal was acquired using Nellcor OxiMax N-600x monitor (Medtronic, Minneapolis, USA). Although the signals were acquired at 1000 sample per second, both BP and PPG signals were down sampled to 100 samples per second. This was done to make computation of the models more efficient, while still preserving the important dynamics of the signal. A typical blood pressure signal and its systolic and diastolic values are shown in Fig. 2. The black line shows the breath holds intervals whenever it jumps up. At its low values, the subject is in normal breathing condition. Obviously, the blood pressure follows a rising trend during BHs and it can be seen in both systolic values interpolated in red and diastolic values interpolated in green.

### C. Model Implementation

Applying ARMA model (Eq. (1)), we obtained the model parameters for estimating SBP (i.e., $\widehat{SBP}$) and DBP (i.e., $\widehat{DBP}$) from PPG signal. ARMA models for each of the five BH intervals were calculated separately for both systolic and diastolic BP in all 7 subjects. In other words, we obtained 5 SBP model and 5 DBP model for each subject. Then, for both measured MBP and estimated MBP (i.e. $\widehat{MAP}$), we applied Eq. (2).

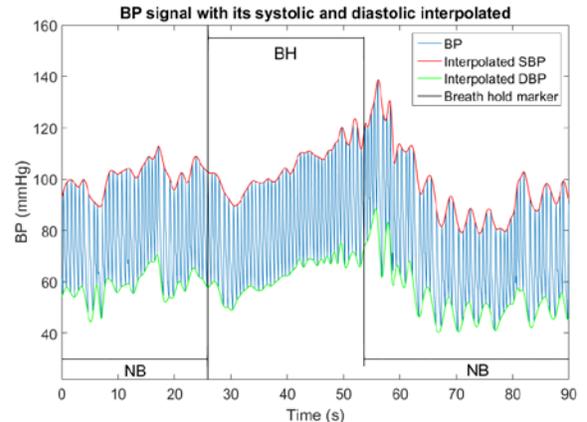

Figure 2- BP signal and systolic and diastolic interpolated values during breath-hold

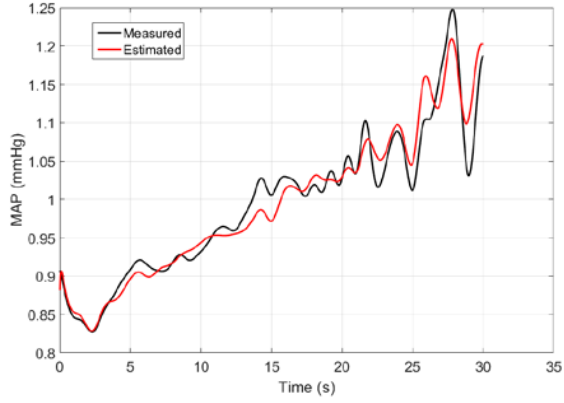

*Figure 3- Estimated MBP signal versus Measured MBP from the same interval (Modeling)*

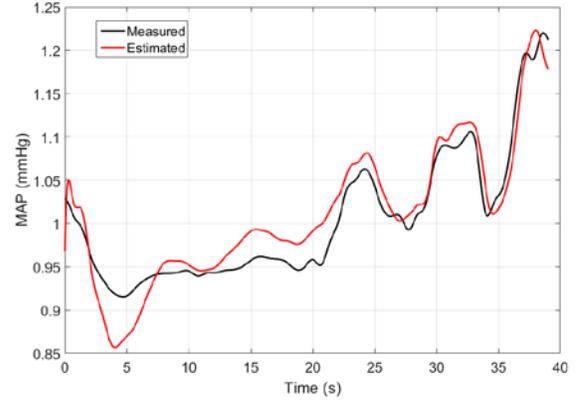

*Figure 4- Estimated MBP signal versus measured MBP in another interval (Validation)*

*D. Model and Cross Validation*

To ascertain the accuracy and predictive ability of the models, we conducted two types of evaluations. First, we computed the residuals for the model by computing the difference between the experimentally measured (e.g., SBP) and the model estimate of the output (e.g. $\hat{SBP}$); referred to as *Model Errors.* Second, we computed the *cross validation errors* by applying a model obtained using one interval to predict the BP in other intervals. We computed the cross validation errors for all unique permutations of models.

## III. RESULTS

Diastolic values are shown in Fig.2. The straight black lines denote the breath holds interval (high level) and normal breathing condition (low level). The rise in blood pressure during BH and its drop afterwards can be easily seen.

Fig. 3 shows a sample of computed $\hat{MAP}$ superimposed on the respective measured values of MBP from the same interval (in this case, BH2). Fig. 4 shows the estimated $\hat{MAP}$ superimposed on measured MBP validated in another interval (in this case, BH2 validated with BH3).

The mean ± standard deviation of the model and cross validation errors for all SBP, DBP, and MBP are shown in Table 1 and Table 2, respectively.

*Table 1- Mean±Std.Dev. of model errors for each BH interval (in mmHg)*

|  | 1st BH | 2nd BH | 3rd BH | 4th BH | 5th BH |
|---|---|---|---|---|---|
| Systolic | 1.00±3.77 | -0.29±3.81 | -2.19±5.73 | -0.88±4.87 | -0.62±4.88 |
| Diastolic | 0.29±2.39 | -0.49±2.63 | -0.66±3.98 | -0.31±3.16 | -0.97±3.63 |
| MBP | 0.53±2.32 | -0.42±2.64 | -1.17±3.82 | -0.50±3.14 | -0.85±3.53 |

*Table 2- Mean±Std.Dev. of cross validation errors of each BH interval obtained using models of all other BH intervals (in mmHg)*

|  | 1st BH | 2nd BH | 3rd BH | 4th BH | 5th BH |
|---|---|---|---|---|---|
| Systolic | 0.28±6.45 | 0.20±5.49 | -1.74±6.55 | -0.93±5.10 | -0.64±4.91 |
| Diastolic | 0.35±4.08 | 0.09±3.37 | -0.44±4.37 | -0.46±3.55 | -0.97±3.67 |
| MBP | 0.33±4.34 | 0.13±3.64 | -0.87±4.42 | -0.62±3.49 | -0.86±3.56 |

To estimate the accuracy of the values of $\hat{SBP}$, $\hat{DBP}$, and $\hat{MAP}$ obtained from each model, the root mean square error (rMSE) for the Model Errors for all BH intervals were computed. The results of averaging the rMSE values derived from the model errors for all subjects are tabulated in Table 3. To assess how a model developed from the data of one interval (e.g. BH1) predicts the corresponding BP measures (i.e. SBP, DBP or MBP) of another congruent interval, we computed the rMSE of the cross validation errors for BH and tabulated them in Table 4.

*Table 3- rMSEs of model errors of BH intervals (in mmHg)*

|  | 1st BH | 2nd BH | 3rd BH | 4th BH | 5th BH |
|---|---|---|---|---|---|
| Systolic | 3.90 | 3.83 | 5.11 | 4.95 | 4.92 |
| Diastolic | 2.41 | 2.68 | 4.03 | 3.18 | 3.76 |
| MBP | 2.38 | 2.67 | 4.00 | 3.18 | 3.63 |

*Table 4- rMSEs of validation of each BH interval by all other BH intervals (in mmHg)*

|  | 1st BH | 2nd BH | 3rd BH | 4th BH | 5th BH |
|---|---|---|---|---|---|
| Systolic | 6.46 | 5.49 | 6.78 | 5.18 | 4.95 |
| Diastolic | 4.10 | 3.37 | 4.39 | 3.58 | 3.80 |
| MBP | 4.35 | 3.65 | 4.50 | 3.54 | 3.66 |

## IV. DISCUSSION

Fig. 3 and Fig. 4 show that $\hat{MAP}$ well track the overall rising trends of the MBP signals during the BH intervals. The means of the model errors and cross validation errors in Table1 and Table 2 provide an assessment of the level of accuracy of estimation of the SBP, DBP, and MBP. As can be seen, the largest estimation errors for BH interval is less than 2.19 mmHg. Indeed, a majority of the cases have error means that are within ±1 mmHg. With taking the level of dispersion of the errors in Table 1 and Table 2 into consideration, it can be seen that the standard deviations for the modeling and validation errors for BH intervals are all below 5 mmHg. This relatively small window of variation is also corroborated by rMSE values that are shown in Table 3 and Table 4.

As can be seen from the results shown in these tables, the upper bound for the dispersion of the cross validation errors is larger than dispersion for model errors (i.e. 6.55 vs 5.73 mmHg, respectively). This is somewhat expected as Cross Validation Errors reflect the ability of the models in predicting BP under the condition different from the condition that model was derived for. In particular, the experimental protocol was designed to examine the effect of successive breath holds on the estimation of BP. In future, we will be reporting on the analysis of NB portion of the experimental protocol.

Comparing the rMSE values in Table 3 with those in Table 4 shows that rMSE values have a max mean of approximately 7 mmHg. Hence, if rMSE is used to gauge the level of the error, for both Model Errors and Cross Validation Errors, an overall error of less than 7 mmHg can be expected. When compared with some of the previously reported techniques, one finds that our results are comparable to techniques which used PTT in estimating BP from PPG signals [15]. These findings show that ARMA model approach is capable of tracking slow frequency trend and also high frequency hemodynamic changes of the body and exhibit adequate accuracy for possible clinical applications.

It is noted, however, that it is unlikely that a single model would be able to estimate BP for members of all population sectors from the PPG variations. The likely need for person-specific models stems from the fact that there is a wide variation in the physiological systems involved in the control of blood pressure which includes responsiveness of the sympathetic nervous system, mechanical, fluid mechanics, dynamical properties of the cardiovascular system, and metabolic rate.

## V. Conclusion

The findings of this pilot study demonstrate that estimating systolic and diastolic BP from PPG measurements using ARMA models can be a viable method for continuous and non-invasive measurement of key BP focal points in OSA subjects with accuracy levels comparable to previously reported values.